\documentclass[12pt,a4paper,floatfix]{iopart}

\usepackage{graphicx}
\usepackage{cite}
\usepackage[breaklinks=true,colorlinks=true,linkcolor=blue,urlcolor=blue,citecolor=blue]{hyperref}

\usepackage{graphicx}
\usepackage{dcolumn}
\usepackage{bm}
\usepackage{epsfig}
\usepackage{amsmath}
\usepackage{amssymb}
\usepackage{xspace}

\begin{document}


\title[The Local Compressibility of Liquids near Non-Adsorbing Substrates]{The Local Compressibility of Liquids near Non-Adsorbing Substrates: A Useful Measure of Solvophobicity {\em and} Hydrophobicity?}

\author{R Evans$^1$ and M C Stewart}
\address{$^1$ H. H. Wills Physics Laboratory, University of Bristol, Bristol BS8 1TL, United Kingdom}
\ead{Bob.Evans@Bristol.ac.uk}

\date{\today}
\begin{abstract}
We investigate the suitability of the local compressibility $\chi(z)$ as a measure of the solvophobicity or hydrophobicity of a substrate. Defining the local compressibility as the derivative of the local one-body density $\rho(z)$ w.r.t. the chemical potential $\mu$ at fixed temperature $T$, we use density functional theory (DFT) to calculate $\chi(z)$ for a model fluid, close to bulk liquid--gas coexistence, at various planar substrates. These range from a `neutral' substrate with a contact angle of $\theta\approx90^{\rm o}$, which favours neither the liquid nor the gas phase, to a very solvophobic, purely repulsive substrate which exhibits complete drying, i.e.\ $\theta=180^{\rm o}$. We find that the maximum in the local compressibility $\chi(z)$, which occurs within one--two molecular diameters of the substrate, and the integrated quantity $\chi_{ex}$ (the surface excess compressibility, defined below) both increase rapidly as $\theta$ increases and the substrate becomes more solvophobic. $\chi(z)$  provides a more pronounced indicator of solvophobicity than the density depletion in the vicinity of the surface which increases only weakly with increasing $\theta$. For the limiting case of drying, $\theta=180^{\rm o}$, we find $\ln\chi(l)\sim l$, where $l$ is the thickness of the intruding film of gas which diverges in the approach to bulk coexistence $\mu\rightarrow\mu_{co}$. When the fluid is confined in a parallel slit with two identical solvophobic walls, or with competing solvophobic and solvophilic walls, $\chi(z)$ close to  the solvophobic wall is altered little from that at the single substrate.  We connect our results with simulation studies of water near to hydrophobic surfaces exploring the relationship between $\chi(z)$ and fluctuations in the local density and between $\chi_{ex}$ and the mean-square fluctuation in the number of adsorbed  molecules.
\end{abstract}

\maketitle

\section{Introduction}
\label{sec:Int}
Understanding, perhaps even defining, hydrophobicity challenges experiment, theory and simulation. Quantifying how water orders near a hydrophobic entity, whether this a small non-polar molecule or a macroscopic substrate such as a Teflon coated cooking pan, is important across many disciplines and many length scales \cite{Chand05,BerWeeksZhou}. From applied physics and materials science perspectives, nanofluidics requires knowledge about slip lengths of water at hydrophobic substrates; these depend upon the structure of the liquid at the substrate which in turn depends upon the substrate--liquid interactions and the commensurability of the liquid with the solid substrate \cite{BocCha}. Hydrophobic interactions are believed to be key-drivers in bio-physical processes such as protein-folding and in micelle and membrane formation \cite{Ball08CR}. Clearly the subject is far-ranging. Here we focus on the nature of adsorption and ordering of liquids at macroscopic substrates. The physical chemist or chemical physicist often poses the question: `How does water order as the degree of hydrophobicity is increased?'  One can, of course, pose the same question for other, simpler, liquids at solvophobic substrates. More precisely one asks: `How does the local ordering of the liquid change as (Young's) contact angle $\theta$, defined for the macroscopic planar substrate, becomes larger?' All physical scientists agree that the larger $\theta$, the less does the substrate prefer the liquid and therefore the more solvophobic/hydrophobic is the substrate. Recall that $\theta$ can take values from $0$ (complete wetting by liquid) to $180^{\rm o}$ which corresponds to complete drying, i.e.\ complete wetting by gas whereby at bulk liquid--gas coexistence a macroscopically thick film of low density gas intrudes between the substrate and the liquid. For contact angles that are large, but $<180^{\rm o}$, answering the question posed is not straightforward and has been addressed from various perspectives. We make no attempt to review the subject fully here. A recent book \cite{BroOle08} provides a useful summary of what is known or surmised from a variety of studies of interfacial and confined water. We also refer the reader to the recent review by Jamadagni {\em et.\ al.} \cite{JamGodGar} entitled `Hydrophobicity of Proteins and Interfaces: Insights from Density Fluctuations' which provides a valuable, interdisciplinary, overview and some motivation for our present study. 

Many treatments of hydrophobicity speak in a descriptive sense about hydrophobic substrates or particles being dewetted \cite{Chand05,BerWeeksZhou} and sometimes invoke a region of depleted water density near the surface as a signature of a hydrophobic substrate. One of the main thrusts of our article is to argue, in keeping with \cite{JamGodGar} and other researchers, that the average one-body density of the liquid/water in the neighbourhood of the substrate might not provide the most effective indicator of the degree of solvophobicity/hydrophobicity. A (hypothetical) {\em very} strongly solvophobic substrate, or a very hydrophobic one, will give rise to a pronounced depletion in local density of the liquid in the immediate neighbourhood of the substrate. However, this situation occurs only for contact angles that are either equal to or very close to $180^{\rm o}$. For substrates realized experimentally, attractive substrate--liquid forces are always present and contact angles are never very close to $180^{\rm o}$ and then one finds the range and magnitude of the depletion are small. Only in the special case of complete drying, which requires a purely repulsive or an extremely weak attractive substrate--liquid interaction, can the range of the depleted (gas) layer become greater than one or two molecular diameters. This statement has not been without controversy. There are many experiments attempting to measure the extent of density depletion often with contradictory results; see the news and views articles \cite{Ball,Chand07} and the summaries of the experimental literature in \cite{BroOle08,JamGodGar}. However, the article by Mezger {\em et\ .al.} \cite{MezReiSchOka} describing high-energy X-ray studies for a water--OTS (octadecyl-trichlorosilane) interface, appears to be definitive. These authors established that there is an integrated density deficit but this corresponds to a small `hydrophobic gap' of about 1.5 water diameters accompanied by an average density reduction (below bulk water) of about 30\%. Simulation studies for realistic water models at substrates with varying degrees of hydrophobicity, as quantified by the contact angle, find similar results. For example, for SPC/E water at non-polar substrates the depletion layer thickness varies from about 1.5 to 2.0 \AA\ as $\theta$ increases in the range $110^{\rm o}$ to $130^{\rm o}$ \cite{JanNet}.

In a recent paper \cite{StewEv14} we speculated that the local compressibility (or local susceptibility) of a {\em simple} liquid, defined for a fixed confining volume, as
\begin{equation}
\chi(z)\equiv\left(\frac{\partial\rho(z)}{\partial\mu}\right)_T,
\label{eq:defchi}
\end{equation}
i.e.\ as the derivative of the local one-body density $\rho(z)$ w.r.t. the chemical potential $\mu$ at fixed temperature $T$, should be much larger in the close vicinity of a solvophobic substrate than it is in the bulk liquid far from the (planar) substrate. Here we employ classical Density Functional Theory (DFT) to calculate $\chi(z)$ for a Lennard-Jones like liquid, close to saturation, at a model solvophobic substrate. We find that $\chi(z)$, obtained for distances $z$ within about two diameters of the substrate, increases rapidly as the contact angle increases. Thus we argue that $\chi(z)$ is a more effective indicator of the degree of solvophobicity than is the local density. Note that $\chi(z)$ was considered in earlier studies of complete drying \cite{EvansParry} and in DFT calculations of gas--liquid interfaces in asymmetrically confined fluids \cite{StewEv12}.

We are not the first to focus on a local compressibility. In the (large) literature on the nature of {\em hydrophobicity} there have been several simulation and theoretical studies that point to the importance of enhanced fluctuations of the local density when water is in contact with a hydrophobic substrate. Chandler and co-workers have written at length on this topic; see e.g.\ Refs. \cite{Chand07,PatVarCha}. However, the subject remains somewhat confused as there does not appear to be a well-accepted measure of the strength and extent of fluctuations---naturally one expects these to be more pronounced as the substrate becomes more hydrophobic/solvophobic. Garde and co-workers come close to considering the $\chi(z)$ that we define above. In a recent paper \cite{AchVemJamGar} this group defines the local compressibility as $\frac{1}{\rho(z)}\left(\frac{\partial\rho(z)}{\partial P}\right)_T$, where $P$ is pressure, and computes this using MD (in the NPT ensemble) for SPC/E water at hydrophobic, self-assembled monolayers (SAMs). They find that this quantity, calculated for $z$ in the immediate vicinity of the most hydrophobic SAMs, can be ten times the corresponding bulk value. As subsequent discussion in Faraday Discussions {\bf 146} reveals, the problem with this definition is that the pressure $P$ is a property of the {\em confined} fluid. In Ref.\ \cite{AchVemJamGar} the pressure $P$  in the derivative was taken as the normal component of the pressure tensor. The same group refers to other measures of a local compressibility \cite{AchVemJamGar,SarGar}. Generally, but not always, the water community is reluctant to take a derivative of the density profile w.r.t. chemical potential -– a natural procedure in an open (grand canonical) treatment of adsorption. Nevertheless, several groups have measured the mean-square fluctuation in particle number for models of water at a hydrophobic substrate or under confinement and argued this becomes large as the degree of hydrophobicity increases. We make contact with some of this work in section \ref{sec:comp} where we discuss the general relationship between the local compressibility $\chi(z)$, as defined above, and the mean-square fluctuation in the number of particles. 

Our paper is arranged as followed: in Sec.\ \ref{sec:comp} we introduce the local compressibility of an adsorbed liquid, defined in (\ref{eq:defchi}), and remind readers how this is related formally to an integral over density-density correlations measured parallel to the substrate. We also introduce the surface excess compressibility and show how this relates to the mean-square fluctuation in the number of adsorbed particles. Sec.\ \ref{sec:model} describes our model liquid and the substrate--liquid effective potentials. We consider the model liquid adsorbed at i) a single substrate (a planar wall) and ii) confined between two planar walls. The DFT that we implement in order to calculate the density profile, contact angle and compressibility is also described in this section. In Sec.\ \ref{sec:res} we present the results of our DFT study and in Sec.~\ref{sec:dis} we comment on how these might be relevant for studies of models of water near hydrophobic substrates. One aim of our paper is to connect the community working on the statistical physics of interfacial phenomena, who naturally consider simple models of liquids, with the chemical physics community who choose to simulate `realistic' models of water and of hydrophobicity. We enquire whether hydrophobicity is inherently different from solvophobicity. The second community often emphasizes the importance of the hydrogen bond network in determining the `unique' structure and anomalous properties of bulk liquid water and argue that when water is close to a hydrophobic substrate the pattern of hydrogen-bonding is disrupted so that the net attractive interactions experienced by the water molecules are reduced compared with bulk. Clearly simple fluids do not exhibit hydrogen bonding. However, at a solvophobic substrate the net attractive interactions experienced by atoms or molecules in a simple fluid are also reduced so one might expect to find a similar variation of the density profiles and of the local compressibility, as a function of contact angle, as for water at a hydrophobic substrate.
\section{ Local compressibility, transverse correlations, fluctuations and sum rules}
\label{sec:comp}

In this section we review briefly some of the formalism of adsorption. We consider a fluid in contact with a reservoir at fixed chemical potential $\mu$ and temperature $T$. For convenience we specialize to either a single planar wall of infinite area that exerts an external potential $V(z)$ on the fluid, or a fluid confined by two parallel planar walls---an open slit pore. In both cases the average one-body density $\rho(z)$ varies only in the $z$ direction normal to the wall(s). It is straightforward to show that the local compressibility defined in (\ref{eq:defchi}) can be expressed \cite{EvansParry,HendvSwol85} as
\begin{equation}
\label{eq:chiG}
\chi(z)=\beta\int^{+\infty}_{-\infty}dz'\int d{\bf R}G(z,z';{R})
\end{equation}
where $\beta=(k_BT)^{-1}$ and $G(z, z'; {R})$ is the density-density pair correlation function for planar geometry; $R=\sqrt{(x-x')^2+(y-y')^2}$ is the transverse separation between particles. The second integral is the zeroth transverse moment of the pair correlation function of the inhomogeneous fluid. For a bulk fluid of uniform density $\rho_b$, Eq.\ (\ref{eq:defchi}) reduces to $\chi_b\equiv(\partial\rho_b/\partial\mu)_T=\rho_b^2\kappa_T$ where $\kappa_T$ is the isothermal compressibility and (\ref{eq:chiG}) reduces to the standard compressibility relation
\begin{equation}
\chi_b=\beta\rho_b\left[\rho_b\int d{\bf r}\left(g(r)-1\right)+1\right]=\beta\rho_bS(k=0),
\label{eq:comp}
\end{equation}
where $g(r)$ is the radial distribution function and $S(k=0)$ is the long-wavelength limit of the liquid structure factor.
 Just as pair correlations become long-ranged in the approach to a bulk critical point, leading to the divergence of $\chi_b$, we expect transverse correlations (parallel to the wall) to become long-ranged in the approach to a surface critical point. For a fluid at a single wall, such behaviour occurs at the critical points of pre-wetting \cite{NicEv} and layering transitions \cite{StewEv14}, both of which correspond to the two-dimensional Ising universality class, and at complete and critical wetting/drying transitions where the details of criticality depend on the nature of the interatomic potentials \cite{DietR}. For a fluid confined by planar walls, separated by a finite distance $L$, there is no longer a wetting or drying transition but pre-wetting and layering can still occur. Moreover capillary condensation and evaporation will occur for the confined fluid. These transitions exhibit criticality that lies in the two-dimensional Ising universality class \cite{EvansParry, EvansMarc87}.

Although in this paper we are not primarily concerned with the divergences of $\chi(z)$, in order to set the scene, it is instructive to consider the behaviour of $\chi(z)$ for a fluid at a single hard-wall with wall--fluid potential
\begin{equation}
V_{hw}(z)=\begin{cases} \infty, \quad  z\leq0
\\
0, \; \: \quad z>0. \end{cases}
\label{eq:Vhw}
\end{equation}
Using the well-known theorem for the density at contact
\begin{equation}
\rho(0^+)=\beta p(\mu)
\label{eq:contact}
\end{equation}
where $p(\mu)$ is the pressure of the reservoir, along with  the Gibbs-Duhem relation, it follows that
\begin{equation}
\chi(0^+)=\beta\rho_b(\mu)\equiv\beta\rho(\infty).
\end{equation}
Thus the compressibility of the fluid at contact is proportional to the density of the fluid {\em far from} the wall. A useful dimensionless measure of the local compressibility is the ratio $\chi(z)/\chi_b$, where $\chi_b$ refers to the fluid in the reservoir. From (\ref{eq:comp}) and (\ref{eq:contact}) we find the ratio of the value at contact to that in the bulk is
\begin{equation}
\frac{\chi(0^+)}{\chi_b}=\frac{1}{S(k=0)}
\label{eq:chicontact}
\end{equation}
where the r.h.s. is evaluated in the bulk (reservoir) fluid.

A particular feature of the hard-wall system is that for simple fluids (Lennard-Jones, square-well etc.) the wall--liquid interface is completely dry, i.e. $\theta=180^{\rm o}$, for all temperatures along the bulk liquid--gas coexistence line $\mu_{co}(T)$. Suppose that the chemical potential is greater than at coexistence so that the bulk fluid far from the wall is a dense liquid. For a Lennard-Jones liquid near its triple point $S(k=0)\sim0.05$. From the measured compressibility of liquid water near its triple point one finds a similar value. In other words the compressibility at contact is about 20 times the value in bulk. Of course, in the limit $\delta\mu\equiv\mu-\mu_{co}\rightarrow0$ a film of gas will intrude at the hard wall and its thickness $l$ will diverge as $l\sim-\ln \delta\mu$ for a three-dimensional fluid in which the interatomic forces are short-ranged. As the gas-liquid interface develops and de-pins from the wall, the local compressibility in this interface also diverges: $\chi(l)\sim\xi^2_{\parallel}\rho'(l)$, where the parallel correlation length $\xi_{\parallel}\sim\delta\mu^{-\nu_{\parallel}}$, with exponent $\nu_{\parallel}=1/2$  for short ranged forces \cite{EvansParry}. This divergence of $\chi(l)$ reflects the development of capillary-wave fluctuations in the de-pinning interface. The same fluctuations lead to a (very weak) broadening of the interfacial part of the density profile so that its derivative w.r.t. $z$ vanishes as $\rho'(l)\sim(-\ln\delta\mu)^{-1/2}$. For $z$ in the close vicinity of the wall the local compressibility is finite and we can estimate $\chi(z\approx0)$ by (\ref{eq:chicontact}).

Here we are concerned with a solvophobic wall where the contact angle $\theta<180^{\rm o}$, so that there is only partial drying. Now there are no divergences as $\delta\mu\rightarrow0^+$. However, this does not mean that the ratio of the local compressibility to the bulk, as defined above, cannot be large. A large local compressibility is a signature of large fluctuations in local density; the correlation length measured parallel to the wall could still be substantial although this does not diverge. Ideally one might attempt to compute the full $G(z, z'; {R})$ and investigate the range of transverse correlations. $\chi(z)$ provides an integrated measure of these at a given distance $z$ from the wall. 

Simulation studies of models of water at hydrophobic substrates have often focused on measurements of the mean-square-fluctuation in the {\em total} number of particles in the simulation system. In order to connect with these studies it is useful to recall some of the thermodynamics of adsorption \cite{EvansMarc87}. A natural quantity to consider is the derivative w.r.t. $\mu$ of the Gibbs excess adsorption $\Gamma_{ex}$, defined as
\begin{equation}
\Gamma_{ex}(\mu)=\int_0^Ldz\left(\rho(z)-\rho_b(\mu)\right)
\label{eq:Gamma_ex}
\end{equation}
where we consider confinement between two planar walls, located at $z=0$ and $z=L$, that exert potentials $V_{w1} (z)$ and $V_{w2}(L-z)$, respectively, on the confined fluid. These external potentials define the excess adsorption, which is the excess amount of fluid per unit area, compared to the uniform bulk fluid occupying the same volume. (In the limit $L\rightarrow\infty$ we recover a single planar wall.) Then we define the surface excess compressibility as
\begin{equation}
\chi_{ex}(\mu)\equiv\left(\frac{\partial \Gamma_{ex}}{\partial \mu}\right)_T=\int_0^Ldz\left(\chi(z)-\chi_b\right),
\label{eq:chi_ex}
\end{equation}
where $L$ is kept fixed. Since this quantity measures the integral of the difference between the local compressibility and the corresponding bulk it provides an integrated measure of `excess density fluctuations' induced by the presence of the wall(s). $\chi_{ex}(\mu)$ diverges at pre-wetting, layering and capillary critical points as the transverse correlations become long ranged but the bulk correlation length, fixed by $(\mu,T)$ in the reservoir, remains finite \cite{EvansMarc87}.
It is straightforward to show\cite{EvansMarc87} that the surface excess compressibility is proportional to the difference between the mean-square fluctuation of the total particle number in the confined fluid and that in the bulk fluid at the same chemical potential:
\begin{equation}
\chi_{ex}(\mu)=\frac{\beta}{A}\left[\left(\langle N^2\rangle-\langle N \rangle^2\right)-\left(\langle N_b^2\rangle-\langle N_b \rangle^2\right)\right],
\label{eq:chi_N}
\end{equation}
and we invoke the thermodynamic limit where the area of the wall $A\rightarrow\infty$. As emphasized in \cite{EvansMarc87} the first term on the r.h.s. must be positive for the confined system to be stable against fluctuations in the total number of particles but since $\chi_{ex}(\mu)$ is the difference between two positive quantities this excess quantity can, in principle, be negative.

In simulation studies on models of confined water what is often measured is the reduced compressibility $\chi_r\equiv\left(\langle N^2\rangle - \langle N\rangle^2\right)/\langle N\rangle$, which in bulk is $S(k=0)$. Clearly this quantity is proportional to the first term in (\ref{eq:chi_N}). Examples of Grand Canonical Monte Carlo (GCMC) simulation results are given in papers by Bratko {\em et.\ al.} \cite{BratCurtBlanPrau,BratDaubLuz,BratDaubLeuLuz,Brat}. An early example is Ref.\ \cite{BratCurtBlanPrau} describing studies of SPC water between two planar apolar walls where capillary evaporation can occur. Others include an illuminating study \cite{BratDaubLeuLuz} of electrostriction of SPC/E water in a model hydrocarbon-like slit pore. The liquid is confined between two hydrophobic Lennard-Jones 9-3 walls corresponding to a contact angle of about $135^{\rm o}$. The authors consider $\chi_r$ for the fluid subject to various applied electric fields. The commentary in Faraday Discussions {\bf 146} \cite{Brat} provides a valuable overview of work from this group.

We note that in \cite{AchVemJamGar} the authors introduce another measure, termed $\chi_{fl}(z)$, of the local compressibility that is based on some measure of the mean-square fluctuation in number of water molecules at a given $z$. However, it is not clear that this is computed at a fixed chemical potential. In \cite{SarGar} one finds results for the water hydration shell compressibility; it is not absolutely clear how this is computed. Other authors, in particular the group of Hummer, have introduced definitions of the isothermal compressibility for a confined fluid that refer to $\left(\partial L/\partial P\right)_T$, where $P$ can be either the normal or parallel component of the pressure tensor of the confined fluid \cite{MitHum}. Another important paper is the GCMC investigation \cite{PertGrun} of TIP4P water confined in an asymmetric slit that has hydrophilic and hydrophobic confining walls. The authors measure the mean-square fluctuation in the total number of molecules for chemical potentials close to bulk coexistence \cite{PertGrun}. In what follows we present results for the density profiles, local compressibility and surface excess compressibility for a simple model liquid subject to solvophobic confining walls.
\section{The Model and the DFT Approach}
 In this section we briefly describe our model fluid, wall potentials and density functional theory methods; these have been described in more detail in a recent paper by the authors \cite{StewEv14}. The choice of parameters in the fluid--fluid and wall--fluid potentials \cite{StewEv14} was motivated by an earlier GCMC study by Pertsin and Grunze \cite{PertGrun} who modelled water confined between hydrophobic and hydrophilic walls.
\label{sec:model}
\subsection{Fluid--fluid potential}
\label{sec:FFP}
In our model fluid, the `molecules' are approximated by hard-spheres with an attractive cut and shifted Lennard-Jones potential between the centres of the spheres:
\begin{equation}
\phi_{att}(r)=
\begin{cases}
-\epsilon - 4\epsilon\left[\left(\frac{\sigma}{r_c}\right)^{12}-\left(\frac{\sigma}{r_c}\right)^6\right] \hspace{5mm} r<r_{min}
\\
4\epsilon\left(\left[\left(\frac{\sigma}{r}\right)^{12}-\left(\frac{\sigma}{r}\right)^6\right]- \left[\left(\frac{\sigma}{r_c}\right)^{12}-\left(\frac{\sigma}{r_c}\right)^6\right]\right)  \hspace{3mm}  r_{min}<r<r_c
\\
0 \hspace{49mm}  r>r_c.
\end{cases}
\label{eq:ffpot}
\end{equation}
As in Ref.\ \cite{StewEv14}, $\sigma=3.154$\AA, $r_{min}=2^{1/6}\sigma$, the cutoff $r_c=2.283\sigma=7.20$\AA\ and $\epsilon=1.55$ kcal/mol ($\epsilon/k_B=780$K). The hard-sphere diameter is taken to be $d=3.00$\AA\ $=0.95\sigma$. Unlike models for water molecules used in simulations, e.g.\ the TIP4P model \cite{JorgChand} used in \cite{PertGrun}, there are no Coulomb interactions between our fluid molecules so the net short-range attraction between molecules is much less than that in simulations. Moreover our model will not exhibit any effects of hydrogen-bonding.

\subsection{Wall--fluid potentials}
\label{sec:WFP}
In this paper we consider four different substrates: a solvophilic wall, a `neutral' wall, a solvophobic wall and a {\em very} solvophobic, purely repulsive wall. The wall potentials for the solvophilic, solvophobic and `neutral' substrates are cut and shifted Lennard-Jones ($9,3$) potentials:
\begin{equation}
V_w(z)=
\begin{cases}
\frac{\epsilon_w}{2}\left[\left(\frac{\zeta}{z}\right)^9-3\left(\frac{\zeta}{z}\right)^3-\left(\frac{\zeta}{z_c}\right)^9+3\left(\frac{\zeta}{z_c}\right)^3\right]  z<z_c
\\
0 \hspace{8mm} z>z_c,
\end{cases}
\label{eq:wpot}
\end{equation}
for a wall positioned at $z=0$, where $\zeta$ is a measure of the range of the potential and we set $\zeta=d$. The cutoff is the same as for the fluid--fluid potential, $z_c=r_c$. As in \cite{StewEv14} the strength of the potential for the solvophilic wall is $\epsilon_w=6.953$ kcal/mol ($\epsilon_w/k_B=3500$K), giving a well depth of $6.2$ kcal/mol. The potential well depth for the `neutral' wall is $3.0$ kcal/mol ($\epsilon_w=3.364$ kcal/mol, $\epsilon_w/k_B=1690$K) and for the weakly attractive, solvophobic wall is $0.46$ kcal/mol ($\epsilon_w=0.516$ kcal/mol, $\epsilon_w/k_B=260$K). The very solvophobic wall has a purely repulsive potential:
\begin{equation}
V_w^{rep}(z)=
\begin{cases}
\frac{\epsilon_w}{2}\left[\left(\frac{\zeta}{z}\right)^9-3\left(\frac{\zeta}{z}\right)^3-\left(\frac{\zeta}{z_c}\right)^9+3\left(\frac{\zeta}{z_c}\right)^3\right] z<z_0
\\
0 \hspace{8mm} z>z_0,
\end{cases}
\label{eq:wpotr}
\end{equation}
where $z_0$ is the point where the wall potential in Eq.\ (\ref{eq:wpot}) crosses the $z$-axis, i.e. $V_w(z_0)=0$. The strength of the potential is $\epsilon_w=0.516$ kcal/mol ($\epsilon_w/k_B=260$K).
\subsection{Density Functional Theory approach}
\label{sec:DFT}

We use density functional theory (DFT) to calculate the equilibrium density profiles of our fluid at the different substrates and hence the local compressibility. In DFT the free energy of an inhomogeneous fluid is expressed as a functional of the average one-body density $\rho(\bf{r})$ (for a review of DFT see Ref.\ \cite{EvR}). The approximation that we employ is a standard one; the excess hard sphere part of the Helmholtz free energy functional ${\cal F}_{ex}^{hs}$ is treated by means of Rosenfeld's fundamental measures theory \cite{Rosen89} and the attractive part of the fluid--fluid interaction potential is treated in mean-field fashion. This approximate functional is the same as that used in Ref.\ \cite{StewEv}, where it is described in more detail. The equilibrium density profile was found by minimising the grand potential functional:
\begin{equation}
\Omega_V[\rho]={\cal F}_{id}[\rho]+{\cal F}_{ex}^{hs}[\rho]+\frac{1}{2}\int{\int{{\rm d}{\mathbf r}_1{\rm d}{\mathbf r}_2\ \rho({\mathbf r}_1)\rho({\mathbf r}_2)\phi_{att}(\lvert{\mathbf r}_1-{\mathbf r}_2\rvert)}} +\int{\rho({\mathbf r})(V({\mathbf r})-\mu)\,{\rm d}{\mathbf r}},
\label{eq:FGP}
\end{equation}
where the density profile $\rho({\mathbf r})=\rho(z)$ for the planar geometry we investigate. The external potential corresponding to the semi-infinite fluid at a single wall is $V({\mathbf r})\equiv V_w(z)$, with $V_w(z)$ given by Eq.\  (\ref{eq:wpot}) for the solvophilic, `neutral' and solvophobic substrates and by Eq.\ (\ref{eq:wpotr}) for the very solvophobic, purely repulsive substrate. For the fluid confined between two parallel walls $V({\mathbf r})\equiv V(z;L)=V_{w1}(z)+V_{w2}(L-z)$, where $V_{w1}$ is the potential of the first wall, positioned at $z=0$ and $V_{w2}$ is the potential of the second wall, positioned at $z=L$. ${\cal F}_{id}[\rho]$ is the Helmholtz free energy functional for the ideal gas. The attractive fluid--fluid potential $\phi_{att}$ is given by Eq.\  (\ref{eq:ffpot}), and with this choice the homogeneous fluid described by Eq.\ (\ref{eq:FGP}) has a (mean-field) critical temperature $k_BT_C/\epsilon=1.35$ and density $\rho_Cd^3=0.2457$.

The local compressibility was calculated from the change in the density profile for a small increase in reservoir chemical potential:
\begin{equation}
\chi(z)=\frac{\rho(\mu+\Delta\mu;z)-\rho(\mu;z)}{\Delta\mu}
\label{eq:chi}
\end{equation}
where the small change in chemical potential was typically chosen to be $\beta\Delta\mu=1\times10^{-10}$.

For each substrate we considered the adsorbed liquid, at $\delta\mu(T)=0^+$, and the adsorbed gas at $\delta\mu(T)=0^-$, and determined the wall--liquid $\gamma_{wl}(\mu_{co})$ and wall--gas $\gamma_{wg}(\mu_{co})$ surface tensions. The contact angle $\theta(T)$ is then obtained via Young's equation:
\begin{equation}
\gamma_{wg}(\mu_{co})=\gamma_{wl}(\mu_{co})+\gamma_{gl}\cos\theta.
\end{equation}
The gas--liquid surface tension $\gamma_{gl}$ was also determined by using DFT to find the density profile and excess grand potential for the gas--liquid interface.
\section{DFT Results}
\label{sec:res}
\subsection{The Local Compressibility for a single wall}
\label{sec:single}
\begin{figure}
\centering
\epsfig{figure=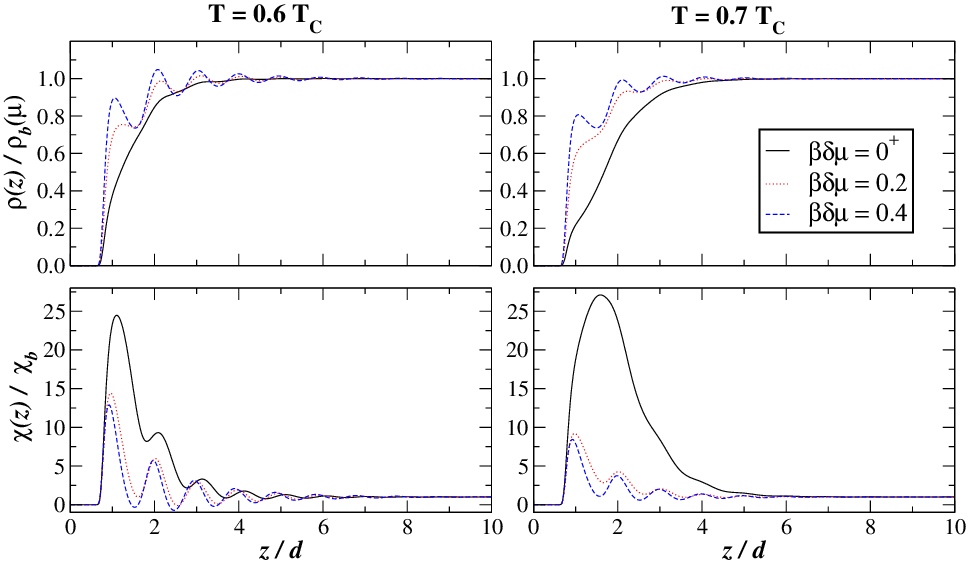, width=13.5cm}
\caption{Density profiles $\rho(z)/\rho_b(\mu)$ (top panels) and local compressibilities $\chi(z)/\chi_b$ (bottom panels), calculated using Eq.\ \ref{eq:chi}), for a fluid at a single solvophobic wall, at three different chemical potentials. The potential well depth for the wall was $0.46$kcal, which results in a contact angle of $\theta=161.9^{\rm o}$ at the temperature $T=0.6T_C$ (left panels) and $\theta=167.7^{\rm o}$ at $T=0.7T_C$ (right panels). The fluid reservoir is on the liquid side of bulk liquid--gas coexistence and the chemical potential deviations of the reservoir from bulk coexistence $\delta\mu=\mu-\mu_{co}$ are given in the legend.}
\label{fig:locsusSolvo0_6_0_7}
\end{figure}

Figure \ref{fig:locsusSolvo0_6_0_7} shows the density profiles and local compressibilities for the fluid at a single solvophobic wall at two different temperatures: $T=0.6T_C$ and $T=0.7T_C$. When the fluid reservoir is at bulk liquid-gas coexistence ($\delta\mu=0^+$) we find a peak in the local compressibility near to this solvophobic wall, where $\chi(z)$  is approximately 25 times the bulk fluid value $\chi_b$. This peak in the local compressibility coincides with the region of depleted fluid density next to the solvophobic wall. The region of depleted density is slightly larger, and the peak in the local compressibility broader, at the higher temperature ($T=0.7T_C$) because the system is closer to a drying transition, as reflected by the increased contact angle; see caption. (Recall that for complete drying the contact angle $\theta=180^{\rm o}$). Away from bulk liquid--gas coexistence, at $\beta\delta\mu=0.2$ and $\beta\delta\mu=0.4$, oscillations develop in the density profile near to the wall and also in the local compressibility. The density next to the solvophobic wall is only slightly less than that in the bulk but the peak in $\chi(z)$ nearest to the wall is around 13 times the bulk value at $T=0.6T_C$ and about 8 times the bulk value at $T=0.7T_C$; at each temperature the maximum in $\chi(z)$ is only slightly higher at $\beta\delta\mu=0.2$ compared to $\beta\delta\mu=0.4$.
\begin{figure}
\centering
\epsfig{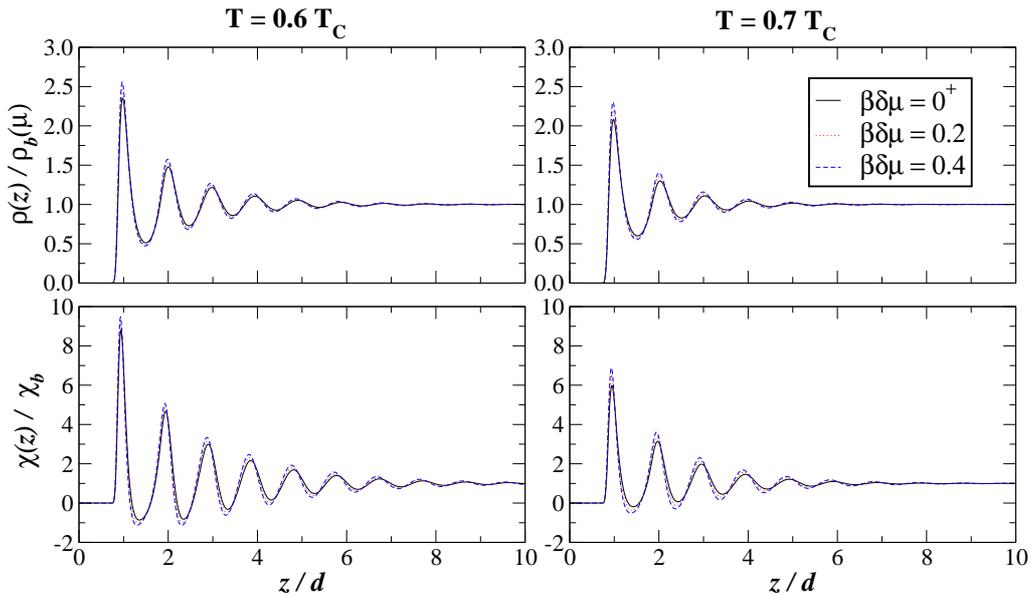}
\caption{Density profiles $\rho(z)/\rho_b(\mu)$ (top panels) and local compressibilities $\chi(z)/\chi_b$  (bottom panels), calculated using Eq.\ \ref{eq:chi}, for a fluid at a single `neutral' wall, at three different chemical potentials. The potential well depth for the wall was $3.0$kcal, which results in a contact angle of $\theta=83.9^{\rm o}$ at the temperature $T=0.6T_C$ (left panels) and $\theta=75.4^{\rm o}$ at $T=0.7T_C$ (right panels). The chemical potential deviations from bulk coexistence $\delta\mu$ are given in the legend.}
\label{fig:locsusNeutral0_6_0_7}
\end{figure}

\begin{figure}
\centering
\epsfig{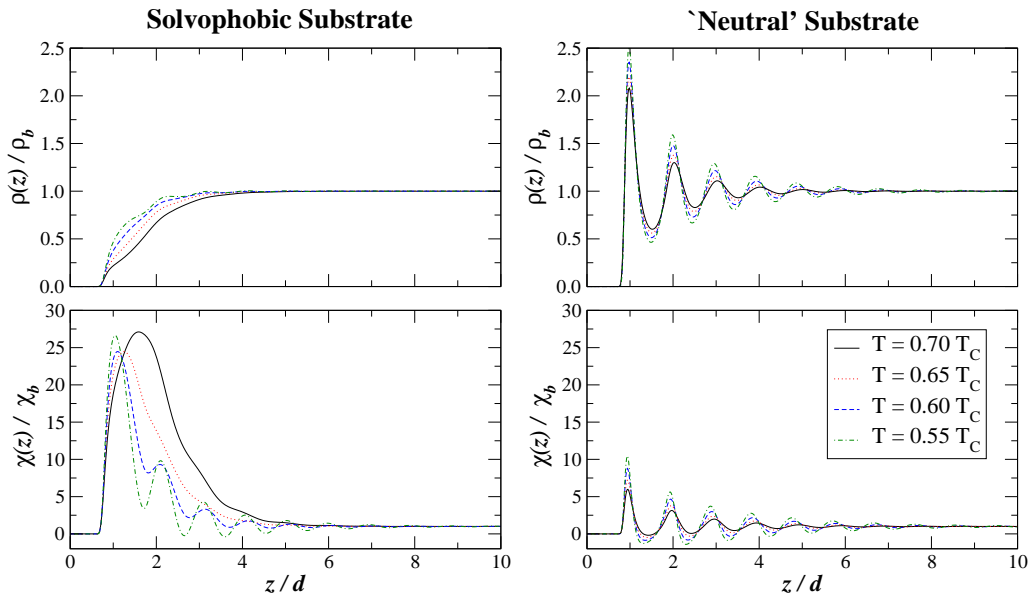}
\caption{Density profiles (top panels) $\rho(z)/\rho_b(\mu)$ and local compressibilities $\chi(z)/\chi_b$ (bottom panels) for a fluid at two different single walls, at different temperatures (see legend) and at bulk liquid-gas coexistence $\delta\mu=0^+$. The potential well depth for the solvophobic wall (left panels) was $0.46$kcal, which results in contact angles of $160.2^{\rm o},161.9^{\rm o},165.4$ and $167.7^{\rm o}$ at the temperatures $0.55T_C, 0.6T_C, 0.65T_C$ and $0.7T_C$, respectively. The potential well depth for the `neutral' wall (right panels) was $3.0$kcal, which results in contact angles of $87.0^{\rm o},83.9^{\rm o},79.8^{\rm o}$ and $75.4^{\rm o}$ at the temperatures $0.55T_C, 0.6T_C, 0.65T_C$ and $0.7T_C$ respectively.}
\label{fig:locsusSolvoNeutralCoexist}
\end{figure}

We now consider a `neutral' wall which has a contact angle close to $\theta=90^{\rm o}$ and therefore favours the liquid and the gas phases approximately equally. Figure \ref{fig:locsusNeutral0_6_0_7} shows the density profiles and local compressibilities for a fluid at this wall at two different temperatures and three different chemical potentials on the liquid side of liquid-gas coexistence. We observe that changing the chemical potential within the range $0^+<\beta\delta\mu<0.4$ has little effect on the density profile and local compressibility at this wall --- both are strongly oscillatory, the oscillations having greater amplitude at the lower temperature $T=0.6T_C$. In Figure \ref{fig:locsusSolvoNeutralCoexist} the density profiles and local compressibilities evaluated at bulk liquid--gas coexistence for the solvophobic and `neutral' wall are shown side by side for comparison. $\chi(z)$ near to the solvophobic wall takes much higher values than at the `neutral' wall throughout the temperature range investigated. At the `neutral' wall the peaks in $\chi(z)$ align with the peaks in the density profile and the oscillations in both the density profile and $\chi(z)$ are more pronounced at lower temperatures. At the solvophobic wall the peak in $\chi(z)$ is clearly associated with the region of depleted fluid density next to the wall and this peak becomes broader as the temperature increases and the contact angle for the fluid at the wall increases. Even at the lowest temperature $T=0.55T_C$, where the fluid density at the wall is only slightly below that of the bulk, the peak in $\chi(z)$ next to the solvophobic wall is much greater and more extended than at the `neutral' wall.

\begin{figure}
\centering
\epsfig{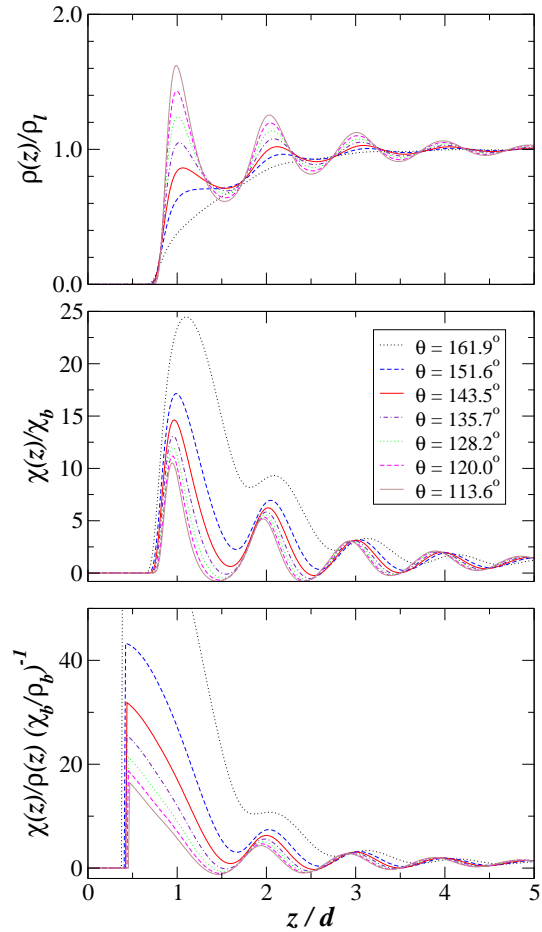}
\caption{Density profiles $\rho(z)/\rho_b(\mu)$ (top panel), local compressibility $\chi(z)/\chi_b$  (middle panel) and local compressibility divided by local density  $\chi(z)/\rho(z) (\chi_b/\rho_b)^{-1}$ (bottom panel) for a fluid at different solvophobic walls, at bulk coexistence $\delta\mu=0^+$ and temperature $T=0.6T_C$. The potential well depths for the walls were 0.46kcal, 0.75kcal, 1.0kcal, 1.25kcal, 1.5kcal 1.75kcal and 2.0kcal; the respective contact angles are given in the legend.}
\label{fig:locsus0_6mu_co}
\end{figure}

Figure \ref{fig:locsus0_6mu_co} shows density profiles and local compressibilities for the liquid at weakly attractive walls ranging from slightly solvophobic, with a contact angle of $113.6^{\rm o}$, to strongly solvophobic with a contact angle of $161.9^{\rm o}$. A typical hydrophobic substrate would lie in this regime. At the strongly solvophobic, i.e.\ only very weakly attractive, end of this range we find that $\rho(z)$ is somewhat depleted in the region one--two molecular diameters from the wall but for less solvophobic walls with $\theta \lesssim 140^{\rm o}$ the local density close to the wall is similar to or greater than the bulk liquid density. $\chi(z)$ exhibits a pronounced maximum close to each of the substrates and the peak height is greater the more solvophobic the wall. In addition to $\chi(z)/\chi_b$ we choose to plot the ratio $\chi(z)/\rho(z)(\chi_b/\rho_b)^{-1}$. We find that this quantity is a maximum for the liquid at contact with the solvophobic wall; the peak is broader and the maximum greater the higher the contact angle. We shall return to this quantity in the discussion where we make a comparison with a recent simulation study of water that covers a range of hydrophobic substrates.

\begin{figure}
\centering
\epsfig{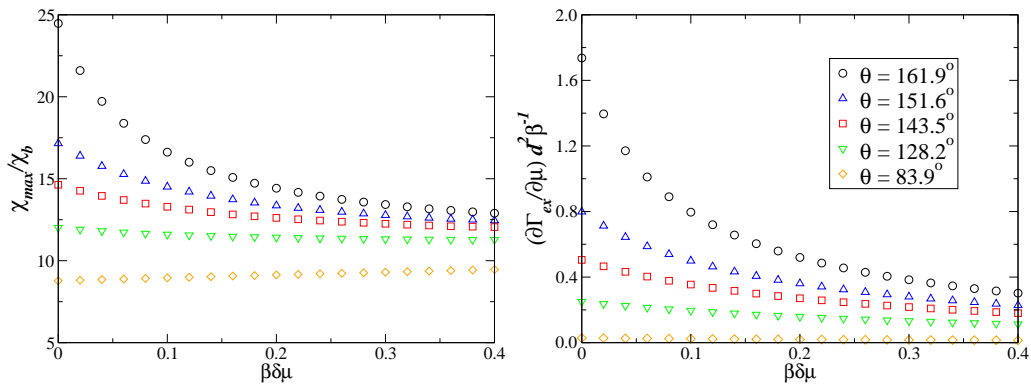}
\caption{The maximum value of the local compressibility $\chi_{max}/\chi_b$ (left panel) and the surface excess compressibility (right panel), calculated from $\chi_{ex}(\mu)=(\partial\Gamma_{ex}/\partial\mu)_T$, versus chemical potential deviation from coexistence $\delta\mu$, for the semi-infinite fluid at five different substrates. The contact angles for the different substrates are given in the legend. The temperature  $T=0.6T_C$. The excess adsorption $\Gamma_{ex}$ was calculated from the density profile using (\ref{eq:Gamma_ex}).}
\label{fig:maxtotsus0_6}
\end{figure}

In Figure \ref{fig:maxtotsus0_6} we consider five different solvophobic walls at a single temperature $T=0.6T_C$. The contact angles of the walls range from $\theta=161.9^{\rm o}$ (this is the solvophobic wall in Figs.\ \ref{fig:locsusSolvo0_6_0_7} and \ref{fig:locsusSolvoNeutralCoexist}) to $\theta=83.9^{\rm o}$ (the neutral wall in Figs.\ \ref{fig:locsusNeutral0_6_0_7} and \ref{fig:locsusSolvoNeutralCoexist}). We plot the maximum value of the relative local compressibility $\chi_{max}/\chi_b$ and the surface excess compressibility $\chi_{ex}$, defined in Eq.\ (\ref{eq:chi_ex}), as functions of the chemical potential deviation from bulk coexistence $\delta\mu$. For all four solvophobic walls, both the maximum compressibilty and the surface excess compressibility are greatest at bulk liquid--gas coexistence and are higher for the  more solvophobic walls. In contrast, the surface excess compressibility for the `neutral' wall is very small and varies only slightly with $\delta\mu$ and the relative maximum compressibility $\chi_{max}/\chi_b$ at the `neutral' wall increases slightly with increasing $\delta\mu$ as a consequence of the bulk quantity $\chi_b$ decreasing.

\begin{figure}
\centering
\epsfig{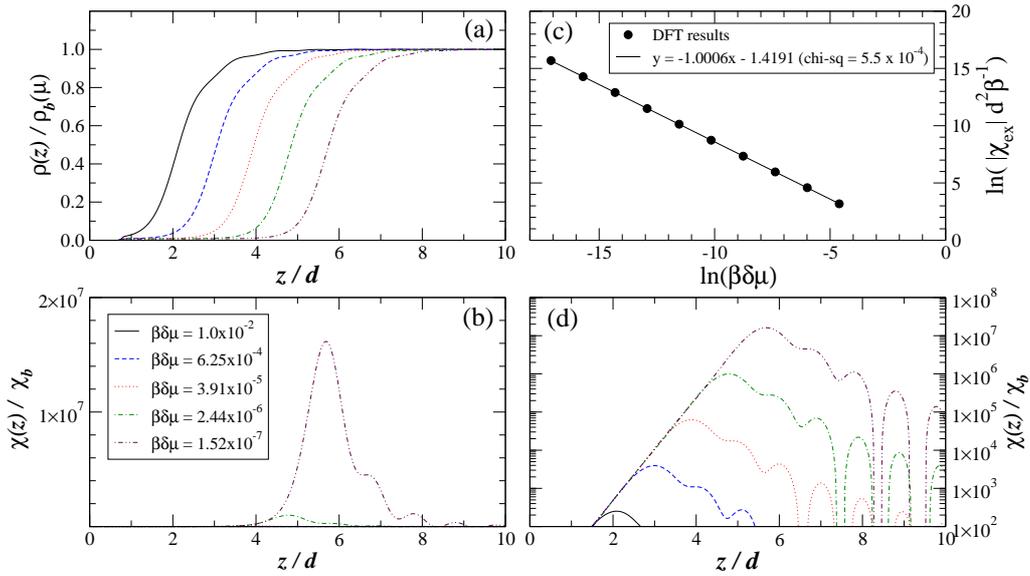}
\caption{Density profiles $\rho(z)/\rho_b(\mu)$ (a) and local compressibility $\chi(z)/\chi_b$ ((b) linear scale, (d) logarithmic scale) for a fluid at a single solvophobic wall, at different chemical potentials approaching bulk liquid--gas coexistence. The wall was purely repulsive and completely dry ($\theta=180^{\rm o}$) at the temperature under investigation,  $T=0.6T_C$. The chemical potential deviations from bulk coexistence $\delta\mu$ are given in the legend. The log--log plot of surface excess compressibility $\chi_{ex}$ versus $\delta\mu$ (c) has a gradient very close to $-1$ confirming that we are observing the growth of a drying film.}
\label{fig:locsusDrying0_6}
\end{figure}

We now study the compressibility in a system that has a contact angle $\theta=180^{\rm o}$ and therefore exhibits complete drying. This is the situation mentioned in Sections \ref{sec:Int} and \ref{sec:comp}: as bulk liquid-gas coexistence is approached from the liquid side, a layer of gas intrudes between the wall and the bulk liquid. The thickness of this layer of gas diverges at bulk coexistence. The wall that we consider is purely repulsive, with wall-fluid potential given by Eq.\ \ref{eq:wpotr}. The density profiles for the liquid at this wall, Figure \ref{fig:locsusDrying0_6}a,  clearly show the growth of the drying film as the chemical potential approaches liquid--gas coexistence $\delta\mu\rightarrow0^{+}$. The local compressibility (Fig.\ \ref{fig:locsusDrying0_6}b,d) is seen to increase very rapidly in the region of the developing gas--liquid interface, with the peak in $\chi(z)$ coinciding with, $l$, the location of the centre of the gas--liquid interface. In Section \ref{sec:comp} we argued that for a system with short-range intermolecular forces exhibiting complete drying the thickness of the drying film diverges as $l\sim -\ln(\delta\mu)$ and therefore the surface excess compressibility diverges as $\partial l/\partial \mu\sim (\delta\mu)^{-1}$. This prediction is confirmed by our DFT results, as illustrated by a log-log plot of the excess compressibility versus $\delta\mu$ (Figure \ref{fig:locsusDrying0_6}c), to which a linear fit gives a gradient very close to $-1$. DFT omits the effects of capillary-wave broadening of the interface so we expect $\chi(l)\sim(\delta\mu)^{-1}$ and therefore $\ln \chi(l)\sim l$. The log plot of $\chi(z)/\chi_b$ in Fig.\ \ref{fig:locsusDrying0_6}d shows that the DFT results are consistent with this prediction.

\subsection{The Local Compressibility of a Confined Fluid}
\label{sec:confined}
\begin{figure}
\centering
\epsfig{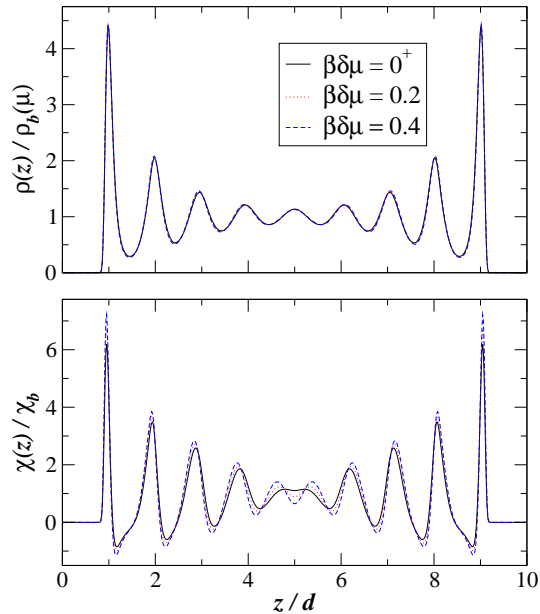}
\caption{Density profiles $\rho(z)/\rho_b(\mu)$ (top panel) and local compressibility $\chi(z)/\chi_b$ (bottom panel) for a fluid confined between two solvophilic walls separated by distance $L=10d$, at three different chemical potentials. The potential well depth for a single wall was $6.2$kcal which results in a contact angle for the semi-infinite fluid at such a wall $\theta=0.4^{\rm{o}}$, at the temperature under investigation,  $T=0.6T_C$. The chemical potential deviations from bulk coexistence $\delta\mu$ are given in the legend.}
\label{fig:locsusTwoPhilic}
\end{figure}

In this sub-section we study the effects of confinement on the local compressibility of a fluid. Figure \ref{fig:locsusTwoPhilic} displays the density profiles and local compressibilities for a fluid confined between two identical parallel solvophilic walls, one at $z=0$ and the other at $z=10d$, where $d$ is the fluid molecular diameter. The walls strongly favour the liquid and have a very small contact angle of $0.4^{\rm o}$ at the temperature of our investigation $T=0.6T_C$. The density profiles are highly oscillatory and the oscillations in $\chi(z)$ closely follow those in the density profile. The local compressibility is highest next to the walls where it is around 7 times the bulk value. There is very little change in the density profile or the local compressibility as the chemical potential is increased from $\beta\delta\mu=0$ to $0.4$; near to the walls they are nearly identical and there is just a small increase in the oscillations of $\chi(z)$ for $z$ close to the centre of the slit at the higher chemical potentials.
\begin{figure}
\centering
\epsfig{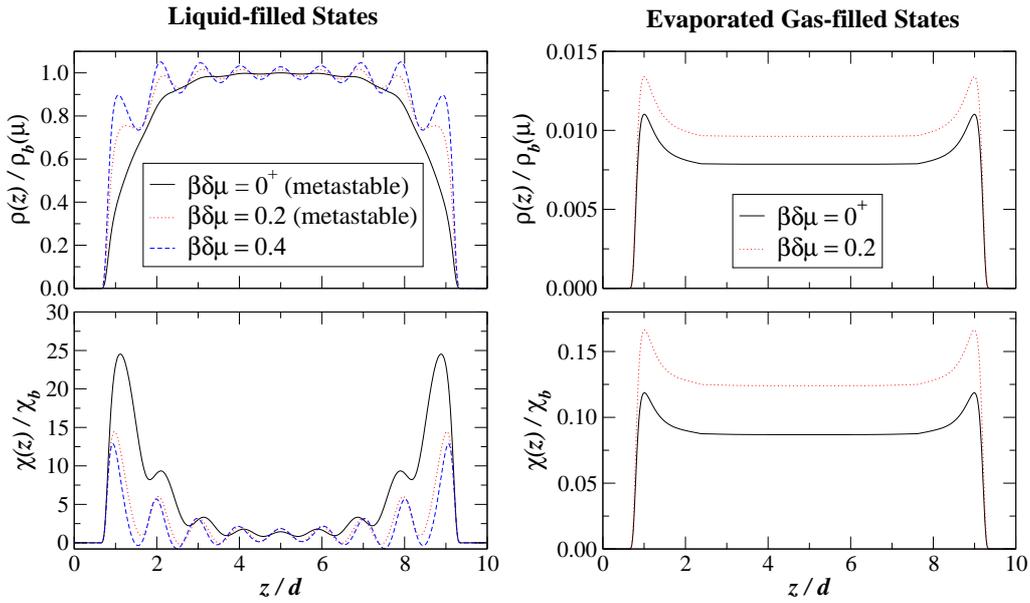}
\caption{Density profiles $\rho(z)/\rho_b(\mu)$ (top panels) and local compressibility $\chi(z)/\chi_b$ (bottom panels) for a fluid confined between two solvophobic walls separated by distance $L=10d$, at three different chemical potentials. The contact angle for the semi-infinite fluid at the single solvophilic wall was $\theta=161.9^{\rm{o}}$ at the temperature under investigation  $T=0.6T_C$. The chemical potential deviations from bulk coexistence $\delta\mu$ are given in the legends. For $\beta\delta\mu<0.263$ the lowest grand potential for the confined fluid is that for the evaporated state (the slit is filled with fluid with density close to that of the bulk coexisting gas). For $\beta\delta\mu=0^+$ and $0.2$ the figure shows both the evaporated states (right panels) and the metastable liquid filled states (left panels---see legend). Note the different vertical scales.}
\label{fig:locsusTwoPhobic}
\end{figure}
We now consider our fluid confined between identical solvophobic walls. When the fluid is near to bulk liquid--gas coexistence capillary evaporation may occur, i.e.\ the slit may be occupied by fluid at density close to that of the bulk gas. We can estimate the chemical potential at which capillary evaporation occurs by comparing the excess grand potential of the evaporated state $\Omega_{ex}^{g}(\mu,L)$ with the excess grand potential of the liquid--filled state $\Omega_{ex}^l(\mu,L)$. The evaporated state is the most stable state when $\Omega_{ex}^g(\mu,L)<\Omega_{ex}^l(\mu,L)$. We can approximate the excess grand potential per unit area in the liquid--filled state to the sum of the surface tensions of the fluid at two separate wall--liquid interfaces $\gamma_{wl}$:
\begin{equation}
\frac{\Omega_{ex}^{l}(\mu,L)}{A}\approx 2\gamma_{wl}(\mu).
\label{eq:gpl}
\end{equation}
Similarly, the excess grand potential per unit area for the evaporated gas filled state is approximated as:
\begin{equation}
\frac{\Omega_{ex}^{g}(\mu,L)}{A}\approx2\gamma_{wg}(\mu)+\delta\mu(\rho_l-\rho_g)L
\label{eq:gpg}
\end{equation}
where $\gamma_{wg}$ is the surface tension of the semi-infinite gas at the wall and the second term in (\ref{eq:gpg}) is the free-energy cost arising from the gas phase being metastable when $\delta\mu>0$. Using the surface tensions at bulk coexistence $\gamma_{wg}(\mu_{co})$ and $\gamma_{wl}(\mu_{co})$ in Eqs. (\ref{eq:gpl}) and (\ref{eq:gpg}) we can estimate the chemical potential, $\delta\mu_{evap}(L,T)$ at which capillary evaporation will occur. For our fluid confined between solvophobic walls, separated by distance $L=10d$, which have a contact angle of $\theta=161.9^{\rm o}$ at $T=0.6T_C$, we calculate $\beta\delta\mu_{evap}\approx0.20$. This result, which corresponds to a Kelvin approximation \cite{EvansMarc87}, is close to the value we find by comparing the excess grand potential for equilibrium density profiles calculated using DFT, $\beta\delta\mu_{evap}=0.26$. Figure \ref{fig:locsusTwoPhobic} displays the density profiles and compressibilities for the fluid at three different chemical potentials confined between the two solvophobic walls. For $\delta\mu<\delta\mu_{evap}$, plots are shown for both the metastable liquid--filled state (left panels) and the evaporated gas--filled state (right panels). As one would expect, knowing that $\chi_b$ in the bulk gas is much smaller (by a factor of about 11) than that in the bulk liquid, the local compressibility in the evaporated states is very much smaller than in the metastable liquid states at the same chemical potential. The local compressibility in the evaporated states is nearly constant throughout the slit with just a small peak next to each wall, whereas the local compressibility in the liquid--filled states has strong peaks near to the solvophobic walls where $\chi(z)$ is around 25 times the bulk value at liquid--gas coexistence $\beta\delta\mu=0^+$ and around 13 times the bulk value at $\beta\delta\mu=0.4$. In the liquid--filled states the density profiles and local compressibilities near to each wall are very similar to those at the single solvophobic wall (left panels, Fig.\ \ref{fig:locsusSolvo0_6_0_7}).

\begin{figure}
\centering
\epsfig{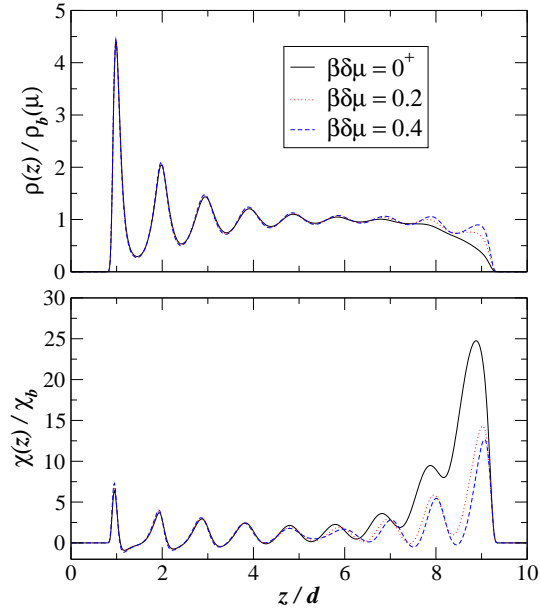}
\caption{Density profiles $\rho(z)/\rho_b(\mu)$ (top panel) and local compressibility $\chi(z)/\chi_b$ (bottom panel) for a fluid confined between a solvophilic wall and a solvophobic wall separated by distance $L=10d$, at three different chemical potentials. The contact angle for the semi-infinite fluid at a single solvophilic wall was $\theta=0.4^{\rm{o}}$ and at the solvophobic wall $\theta=161.9^{\rm{o}}$, at the temperature under investigation,  $T=0.6T_C$. The chemical potential deviations from bulk coexistence $\delta\mu$ are given in the legend.}
\label{fig:locsusTwoAsym}
\end{figure}
Figure \ref{fig:locsusTwoAsym} shows density profiles and local compressibilities for the fluid confined between parallel {\em asymmetric} walls: the left--hand wall is solvophilic and the right--hand wall is solvophobic. The effect of the confinement on the density profiles and the local compressibilites near to each wall is minimal. Near to the solvophilic wall these functions are very similar to those of the semi-infinite fluid at the single solvophilic wall (not shown here) and at the solvophilic walls in the symmetric solvophilic slit (Fig.\ \ref{fig:locsusTwoPhilic}). Near to the solvophobic wall, the large peak in the local compressibility appears to be unaffected by the presence of the solvophilic wall opposite; this is similar to that observed at the single solvophobic wall (Fig.\ \ref{fig:locsusSolvo0_6_0_7}) and at each wall in the symmetric solvophobic slit (Fig.\ \ref{fig:locsusTwoPhobic}). The arrangement of asymmetric walls in Figure \ref{fig:locsusTwoAsym} also gives rise to a metastable state (not shown here) which has a layer of fluid with the gas density next to the solvophobic wall. The grand potential of this state is significantly higher than that of the liquid--filled state. This metastable state is connected to the layering transitions which occur in the semi-infinite fluid at the solvophilic wall (see system (ii) in Ref.\ \cite{StewEv14}).

\section{Summary and Discussion}
\label{sec:dis}
Using classical DFT we have calculated the local compressibility $\chi(z)$ and surface excess compressibility $\chi_{ex}$ for a model liquid near to solvophobic substrates. A significant peak in $\chi(z)$ is observed for a distance $z$ in the vicinity of a solvophobic wall. The height of this peak is greatest for the fluid at bulk liquid-gas coexistence $\delta\mu =0^{+}$ but the peak persists into the region of the bulk phase diagram where the liquid is the stable state, $\delta\mu>0$; see Fig.\ \ref{fig:locsusSolvo0_6_0_7}. The maximum in the local compressibility $\chi_{max}$ and the surface excess compressibility $\chi_{ex}$ increase as the substrate becomes more solvophobic, i.e. for increasing values of the contact angle $\theta$; see Figs.\ \ref{fig:locsusSolvoNeutralCoexist}, \ref{fig:locsus0_6mu_co} and \ref{fig:maxtotsus0_6}. The effect of substrate solvophobicity on $\chi(z)$ is much more pronounced than the effect on the local density $\rho(z)$ of the fluid. $\rho(z)$ is reduced slightly compared to the bulk fluid density $\rho_b$ in the region one--two molecular diameters from the weakly attractive, solvophobic wall. However, the extent of the density depletion does not change vastly with increasing $\theta$. It is only when the wall is made very solvophobic (purely repulsive) that a significant region of low density fluid develops at the wall. This type of wall has a contact angle of $180^{\rm o}$ and exhibits complete drying. Our DFT results for this situation (Fig.\ \ref{fig:locsusDrying0_6}) confirm the predicted divergence of the surface excess compressibility $\chi_{ex}\sim(\delta\mu)^{-1}$ and show that the logarithm of the local compressibility $\ln \chi(l) \sim l$, where the thickness of the drying film $l\sim-\ln (\delta\mu)$, as $\delta\mu\rightarrow0^{+}$. Of course a purely repulsive substrate cannot be realised experimentally -- attractive intermolecular forces will always be present and $\theta$ will be $<180^{\rm{o}}$. In contrast to the solvophobic cases, for a `neutral' wall where $\theta\approx 90^{\rm {o}}$,  increasing $\delta\mu$ has little effect on $\chi(z)$ (Fig.\ \ref{fig:locsusNeutral0_6_0_7}) . Moreover the ratio $\chi_{max}/\chi_b$  and $\chi_{ex}$  are nearly independent of chemical potential and the latter is close to zero (Fig.\ \ref{fig:maxtotsus0_6}).

Confining the liquid between two walls appears to have little effect on $\chi(z)$, i.e.\ near to the walls our results for $\chi(z)$ are very similar to those at the individual single walls; see Figs.\ \ref{fig:locsusTwoPhobic} and \ref{fig:locsusTwoAsym}. Where the two confining walls are both solvophobic, capillary evaporation may occur as bulk liquid--gas coexistence is approached. The slit is occupied by fluid at density close to that of the bulk gas and, as one expects, we find that in this (evaporated) state $\chi(z)/\chi_b$ is very much lower than in the liquid--filled state.

 It is clear from our DFT study of simple fluids at solvophobic substrates that $\chi(z)$ is a useful measure of the degree of solvophobicity. There are pronounced fluctuations in the density for distances close to such a substrate and these grow with decreasing $\delta{\mu}$ and with increasing contact angle. Thus from Sec.\ \ref{sec:comp} we would conclude that transverse correlations, for $z$ and $z'$ within a couple of molecular diameters of the wall, increase in range as the contact angle increases. We emphasize that it is important to distinguish between partial drying, where $\theta<180^{\rm o}$ and there is no diverging transverse correlation length, and the limiting case of complete drying, $\theta=180^{\rm o}$, where the transverse correlation length in the emerging gas--liquid interface does diverge as $\delta{\mu}\rightarrow0$. The distinction is well-known to the statistical physics community \cite{EvansParry,HendvSwol85,DietR,EvR} working on adsorption and wetting but appears to be less known to the water community. Of course the theoretical approach we employ here is a mean-field DFT. As discussed at length in our earlier papers \cite{StewEv14,StewEv12,StewEv,StewEvJPCM}, this approach omits {\em some} of the effects of fluctuations. These will act to smear the density profiles. Although it is likely that the DFT generates density profiles for $\theta<180^{\rm o}$ that are over-structured, i.e. the oscillations of the density could be over-exaggerated and this might lead to oscillations in $\chi(z)$ more pronounced than in simulation, we are confident that the DFT predictions for the overall variation of $\chi(z)$ with contact angle are robust. In this regime, where there is no diverging transverse correlation length, there is no reason to distrust the qualitative, perhaps even the quantitative, predictions of DFT. There is no `free' gas--liquid interface with its accompanying capillary wave-like fluctuations. The interface can only emerge in the extreme case of complete drying, which occurs only for repulsive or very weakly attractive substrates. Examples of how large $\chi(z)$ can be in the presence of a nearly free interface are given in Fig.11 of \cite{StewEv12} (for an asymmetrically confined fluid) and in Fig.\ \ref{fig:locsusDrying0_6} of the present paper (for a single, drying wall). In the latter case, as the interface depins from the substrate, $\chi(l)\sim\delta{\mu}^{-1}$. This is the mean-field result. As mentioned in Sec.\ \ref{sec:comp} including effects of capillary wave broadening would lead to a slightly weaker divergence: $\chi(l)\sim(-\ln\delta{\mu})^{-1/2} \delta{\mu}^{-1}$. Moreover the exponent describing the divergence of the integrated quantity $\chi_{ex}$ is unaffected. Thus we expect $\chi_{ex}\sim\delta{\mu}^{-1}$ when fluctuations are included; only the amplitude should be changed from the mean-field result \cite{EvansParry,DietR}.

In this paper we do not consider the case of critical drying, i.e. the continuous transition $\cos\theta\rightarrow-1$ as the temperature is increased along the bulk coexistence curve \cite{DietR}. Ref.\ \cite{StewEvJPCM} describes a DFT treatment of such a transition (there induced by increasing the strength of wall-fluid attraction) for a model fluid that incorporates long-ranged dispersion forces.

Turning to the relevance of our study for water at hydrophobic substrates, we remark that it should be straightforward to measure the local compressibility $\chi(z)$ and the surface excess $\chi_{ex}$ in GCMC simulations. It would be interesting to examine whether variations with $\delta{\mu}$ and with increasing $\theta$ similar to those reported here are observed. Although we do not claim to have mastered the (vast) literature on simulation of water models at single or confining hydrophobic substrates we have not found results for $\chi(z)$. In Sec.\ \ref{sec:comp} we referred to GCMC studies by Bratko {\em et.\ al.} \cite{BratDaubLeuLuz} for SPC/E water confined between identical Lennard-Jones 9-3 walls, corresponding to a contact angle of about 135$^{\rm o}$, modelling a hydrocarbon-like nanopore. In the absence of an applied electric field the reduced compressibility $\chi_{r} =
(\langle N^2\rangle-\langle N\rangle^2)/\langle N\rangle$, for wall separation $L=2.7$nm, was about three times the value for the bulk liquid at the same chemical potential. In common with other studies, Bratko {\em et.\ al.} \cite{BratDaubLeuLuz} attribute the high value of $\chi_{r}$ to increased fluctuations in number density in the vicinity of the hydrophobic walls. In a subsequent commentary \cite{Brat} Bratko collects GCMC results for $\chi_r$, measured for a slit pore with $L=1.64$nm, and plots these as a function of the contact angle, $\theta$. For $\theta$ below about 90$^{\rm o}$, $\chi_r$ takes values similar to that of the bulk liquid whereas in the range $90^{\rm o}<\theta<135^{\rm o}$, $\chi_r$ increases rapidly with increasing hydrophobicity; see Fig.\ 2 in Ref.\ \cite{Brat}. Since $\chi_r$ measures the mean square fluctuations in the {\em total} number of molecules this quantity need not exhibit the same behaviour as the surface excess compressibility $\chi_{ex}$ that we calculate; see (\ref{eq:chi_N}). Nevertheless, the results we present in Fig.\ \ref{fig:maxtotsus0_6} (right panel), for a single wall with $\delta{\mu}=0^+$, follow a similar trend with $\theta$. For $\theta\lesssim90^{\rm o}$ the excess quantity is close to zero but increases rapidly with increasing $\theta$. It would be interesting to compare simulation results for $\chi_{ex}$ for water at a single hydrophobic substrate with the present ones. 
 Note that Bratko {\em et.\ al.} do not present results for increasing chemical potential; they fix $\delta\mu$ at a constant value. The GCMC study by Pertsin and Grunke \cite{PertGrun} of TIP4P water confined between hydrophilic and hydrophobic walls does measure the variation of $\chi_r$ with chemical potential. They find `giant fluctuations in the number of water molecules'. However the results for $\chi_r$ in their Fig.\ 9 are for the case where the hydrophobic wall is purely repulsive (dry) so that a wandering liquid--gas interface can develop and layering transitions may occur in the confined system \cite{StewEv14}.

It is not appropriate to attempt to review the large number of simulations of water at hydrophobic substrates carried out in other ensembles. We mentioned some of these in Sec.\ \ref{sec:comp} and many references are given in the review \cite{JamGodGar}. Whilst many studies point to enhanced density fluctuations in the neighbourhood of a hydrophobic substrate the authors are not always careful to distinguish between partial ($\theta<180^{\rm o}$) and complete drying ($\theta =180^{\rm o}$). The papers \cite{JamGodGar,AchVemJamGar,SarGar} from Garde's group are in a similar spirit to our present study. As we stated in Sec.\ \ref{sec:Int}, in Ref.\ \cite{AchVemJamGar} results for a local compressibility, closely analogous to $\chi(z)$ but with the derivative taken w.r.t. the normal pressure rather than $\mu$, were presented for SPC/E water at hydrophobic SAMs. The results in their Fig.\ 2 show that the (authors') local compressibility has a maximum in the vicinity of the substrate and that the height of this maximum increases with increasing hydrophobicity.

Clearly the phenomenology that emerges from the simulations of Garde {\em et.\ al.}, and from studies by other groups, is close to that we have ascertained for a simple model fluid at a solvophobic substrate. We would argue that performing GCMC simulations to calculate $\chi(z)$, and $\chi_{ex}$, would  provide cleaner measures of the strength and extent of fluctuations in the local density.  Are there other structural indicators of solvophobicity and hydrophobicity? A large local compressibility signals growth of correlations parallel to the wall. Thus for $z$, $z'$ close to a solvophobic or hydrophobic wall we might expect $G(z,z';{R})$ to exhibit longer ranged decay, with $R$, than the pair correlation function of the bulk fluid at the same chemical potential. Such behaviour is shown in Fig.5 of the review \cite{JamGodGar} where transverse water--water correlations are plotted for a model of water at a 9-3 wall. As the strength of the wall--water attraction is reduced, and the wall becomes more hydrophobic, the range of the transverse correlations increases. As mentioned earlier, we expect to observe the same trend in our model system. In this context a useful quantity to consider is the local structure factor (sometimes called the transverse structure factor \cite{TarEv82}):
\begin{equation}
S(z;q)\equiv \int_{-\infty}^{\infty}dz'G(z,z';q),
\label{eq:Szq}
\end{equation}
where we have taken a Fourier transform w.r.t. ${\bf R}$, i.e.\ $q$ is the transverse wavenumber. From (\ref{eq:chiG}) we see that 
\begin{equation}
\beta^{-1}\chi(z)=S(z;0),
\label{eq:chi_Sz0}
\end{equation}
the long wavelength limit of the local structure factor. $S(z;q)$ provides information about the range (correlation length) and strength of transverse correlations and can be obtained without huge computational effort from DFT \cite{TarEv82}{\footnote{This paper and \cite{FoiAsh} were concerned with the growth of transverse correlations in the approach to wetting.}}.

After completing our calculations we learnt of a very recent molecular dynamics study of SPC/E water at a planar Lennard-Jones 12-6 wall by Willard and Chandler \cite{WillChand}. In common with our density profiles in Fig.\ \ref{fig:locsus0_6mu_co} the authors observe oscillatory density profiles at weakly hydrophobic walls, with the oscillations becoming less pronounced as the wall is made less attractive. A region of depleted fluid density, extending about 3{\AA}, is seen at their most hydrophobic wall; see Fig.\ 2 `standard interface', top panel in Ref.\  \cite{WillChand}. In Ref.\ \cite{WillChand} these changes in the behaviour of the density profile appear to occur over a surprisingly narrow range of contact angle: $120^{\rm o}\lesssim \theta \lesssim133^{\rm o}$ whereas we observe similar changes in our DFT density profiles in the range $130^{\rm o} \lesssim \theta \lesssim 160^{\rm o}$, see Fig.\ \ref{fig:locsus0_6mu_co}. In Ref.\ \cite{WillChand} results are presented for number density fluctuations at hydrophobic walls. The measure the authors adopt is the mean-square fluctuation in $N(z)$, the number of water molecules in a spherical probe volume with radius 3{\AA} centred at distance $z$ from the wall. Results for $(\langle N(z)^2 \rangle -\langle N(z) \rangle ^2)/\langle N(z) \rangle$ for the `standard interface' are plotted in the top panel of their Fig.\ 3. In common with our results in Fig.\ \ref{fig:locsus0_6mu_co} for the ratio $\chi(z)/\rho(z)$, their measure of the density fluctuations peaks for the liquid in contact with the wall and this peak becomes broader and higher for weaker wall--water attraction, i.e.\ increasing hydrophobicity. Although the measures employed in the present paper and in \cite{WillChand} are not identical they are similar and it is striking that the two sets of results display similar features and similar variation with contact angle. Willard and Chandler \cite{WillChand} distinguish between the `intrinsic' and the `standard' interface. The latter uses the usual equilibrium statistical mechanics definition of the one-body density and related quantities. The authors argue that the density profiles, and (their measure of) the density fluctuations, calculated according to their definition of the `intrinsic' interface are unaffected by increasing the strength of wall--water attraction thereby reducing the contact angle. Moreover they find these quantities are close to those obtained for the `intrinsic' {\em free} interface between liquid water and its gas. We find this result curious. As remarked previously---for $\theta$ well below $180^{\rm o}$, and therefore away from the complications of critical drying, the (`standard') gas-liquid  interface is not manifest at the (`standard') solvophobic wall--liquid interface. The latter is in no sense a composite of wall--gas and gas--liquid interfaces. For the largest contact angle in Fig.\ \ref{fig:locsus0_6mu_co}, $\theta =162^{\rm o}$, the density profile resembles a {\em portion} of the gas--liquid profile, c.f.\ the profiles for complete drying in Fig.\ \ref{fig:locsusDrying0_6}a. However, for $\theta \lesssim 130^{\rm o}$, typical of real hydrophobic substrates, the (`standard') wall--liquid density profile and the associated mean-square fluctuation in $N(z)$ bear no resemblance to those of the gas--liquid interface. In this regime it is difficult to imagine that the (`standard') wall--liquid interface `knows' about the incipient gas--liquid interface. Recall that in this regime the surface tension $\gamma_{wl}(\mu_{co})$ is considerably smaller than $\gamma_{wg}(\mu_{co})+\gamma_{gl}$. Evidently the `intrinsic' interface is a very different beast from the `standard' interface and one might enquire whether the former is useful in interpreting simulation/theory results for solvophobic/hydrophobic walls.

We conclude by returning to the question as to whether solvophobicity is inherently different from hydrophobicity. The short answer is probably `No'. Of course chemistry determines whether a particular substrate, say a SAMs hydrocarbon or a paraffin, dislikes water but once one accepts that the net attractive interactions experienced by the water molecules are in some sense reduced compared with those in the bulk liquid then the physics of partial drying, with accompanying density fluctuations, should be essentially the same as for a simple liquid.

\ack
We are grateful to A.J. Archer, D. Chandler and R. Roth for helpful comments. R.E. acknowledges support from the Leverhulme Trust under award EM/2011-080.

\section*{References}

\providecommand{\newblock}{}

\bibliography{thesis_bibfile}

\end{document}